\documentclass[aps,prd,onecolumn,amsfonts,amssymb,amsmath,showpacs,floatfix,nofootinbib,groupedaddress,superscriptaddress,citesort]{revtex4}
\usepackage{mathrsfs}
\usepackage{amsfonts}
\usepackage{amstext}
\usepackage{amsmath}
\usepackage{amssymb}
\usepackage{float}
\usepackage[colorlinks, citecolor=blue]{hyperref}
\usepackage[usenames]{xcolor}
\usepackage[dvips]{graphicx}
\def\qed{\leavevmode\unskip\penalty9999 \hbox{}\nobreak\hfill
     \quad\hbox{\leavevmode  \hbox to.77778em{%
              \hfil\vrule   \vbox to.675em%
               {\hrule width.6em\vfil\hrule}\vrule\hfil}}
     \par\vskip3pt}

\def\ra{\rangle}
\def\la{\langle}

\def\no{\nonumber}
\def\bea{\begin{eqnarray}}
\def\eea{\end{eqnarray}}
\def\be{\begin{equation}}
\def\ee{\end{equation}}
\def\c{\cos r}
\def\s{\sin r}

\def\ctt{\cos^2\frac{\theta}{2}}
\def\stt{\sin^2\frac{\theta}{2}}
\def\cc{\cos^2 r}
\def\ss{\sin^2 r}

\begin{document}
\title{Fine-grained uncertainty relation under the relativistic motion}

\author{Jun Feng}
\email{j.feng@xjtu.edu.cn}
\affiliation{School of Science, Xi'an Jiaotong University, Xi'an 710049, China}
\affiliation{School of Mathematics and Physics, The University of Queensland, Brisbane, QLD 4072, Australia}
\author{Yao-Zhong Zhang}
\affiliation{School of Mathematics and Physics, The University of Queensland, Brisbane, QLD 4072, Australia}
\affiliation{Institute of Modern Physics, Northwest University, Xian 710069, China}
\author{Mark D. Gould}
\affiliation{School of Mathematics and Physics, The University of Queensland, Brisbane, QLD 4072, Australia}
\author{Heng Fan}
\affiliation{Beijing National Laboratory for Condensed Matter Physics,
Institute of Physics, Chinese Academy of Sciences, Beijing 100190, P. R. China }

\begin{abstract}
Among various uncertainty relations, the profound fine-grained uncertainty relation is used to distinguish the uncertainty inherent in obtaining any combination of outcomes for different measurements. In this Letter, we explore this uncertainty relation in relativistic regime. For observer undergoes an uniform acceleration who is immersed in an Unruh thermal bath, we show that the uncertainty bound is dependent on the acceleration parameter and choice of Unruh modes. We find that the measurements in mutually unbiased bases, sharing same uncertainty bound in inertial frame, could be distinguished from each other for a noninertial observer. In an alternative scenario, for the observer restricted in a single rigid cavity, we show that the uncertainty bound exhibits a periodic evolution w.r.t. the duration of acceleration. With properly chosen cavity parameters, the uncertainty bounds could be protected. Moreover, we find that uncertainty bound can be degraded for specific quantum measurements to violate the bound exhibited in nonrelativistic limit, which can be attributed to the entanglement generation between cavity modes during particular epoch. Several implications of our results are discussed.\\

\end{abstract}
\pacs{03.65.Ta, 03.67.Mn, 04.62.+v}
\maketitle

\section{Introduction}
\label{1}


The distinguishability of a quantum theory from its classical counterpart is formulated in Heisenberg uncertainty principle \cite{H1}, which bounds our prediction ability for a quantum system. In terms of entropic measures, it can be recast as \cite{H5} $H(Q)+H(R)\geqslant\log_2\frac{1}{c}$, where $H(Q)$ and $H(R)$ are the Shannon entropy for the probability distribution of measurement outcomes. Since the complementarity $c$ between observables $Q$ and $R$ does not depend on specific states to be measured, the r.h.s. of the inequality provides a fixed lower bound and a more general framework of quantifying uncertainty than standard deviations \cite{H2,H3}. Moreover, once using quantum memory to store information about the measured system, the entropic uncertainty bound could even be violated \cite{EUR2}, due to the entanglement between quantum memory and system. Such entropic uncertainty relation (EUR) plays an important role in many quantum information processes, e.g., quantum key distribution.

Nevertheless, entropic function is still a rather coarse way of measuring the uncertainty of a set of measurements  (see Ref. \cite{REV} for a recent review). For instance, with EUR, one can not distinguish
the uncertainty inherent in obtaining any combination of outcomes for different measurements.
To overcome this defect, a new form of uncertainty relation, i.e., fine-grained uncertainty relation (FGUR), has been proposed recently \cite{FGUR}. For a set of measurements labeled by $t$, associating with every combination of possible outcomes $\textbf{x}=(x^{(1)},\ldots,x^{(n)})$, there exist a set of inequalities
\begin{equation}
\label{FGUR}
   \left\{\sum_{t=1}^{n}p(t)p(x^{(t)}|\rho)\leq\zeta_{\textbf{x}}\Big|\textbf{x}\in\mathbf{B}^{\times n} \right\},
\end{equation}
where $\mathbf{B}^{\times n}$ is the set involving all possible combinations of outcomes, $p(t)$ is the probability of choosing a particular measurement, and $p(x^{(t)}|\rho)$ is the
probability that one obtains the outcome $x^{(t)}$ after performing measurement $t$ on the state
$\rho$. To measure the uncertainty, the maximum in function $\zeta_{\textbf{x}}=\max_{\rho}\sum_{t=1}^n p(t)p(x^{(t)}|\rho)$ should be evaluated over all states allowed on a particular system. It can be proved that one can not obtain outcomes with certainty for all measurements simultaneously when $\zeta_{\textbf{x}}<1$.  Once inequality (\ref{FGUR}) is saturated, state $\rho$ is recognized as maximally certain state (MCS). 

Since its introduction, many applications have been found for the FGUR. For instance, it was shown \cite{FGUR1} that the FGUR could be used to discriminate among classical, quantum, and superquantum correlations involving two or three parties. Moreover, the uncertainty bound in (\ref{FGUR}) could be optimized once the measured system is assisted by a quantum memory \cite{FGUR2}. Moreover, a profound link between the FGUR and the second law of thermodynamics has been found \cite{FGUR3}, which claims that a violation of uncertainty relation implies a violation of thermodynamical law. Other studies from various perspectives could be found in \cite{FGUR4,FGUR5}.

While most of studies on uncertainty relations are nonrelativistic, a complete account of these relations requires one to understand them in relativistic regime, which would link many different physical branches, e.g., quantum information, relativity, and may even shed new light on quantum gravity \cite{APP2}.  In the previous work, we have shown that, besides the choice on the observables, the entropic uncertainty bound should also depend on relativistic motion status of the observer who performs the measurement \cite{FENG2}, or the global structure of curved spacetime background \cite{FENG3}. This new character of quantum-memory-assisted EUR is a direct result of entanglement generation in a relativistic system.

In this Letter, we explore FGUR for quantum system under relativistic motion, and find the uncertainty bound does depend on the motion state of the system. We first consider observer undergoes an \emph{uniform} acceleration relative to an inertial reference. Since two frames differ in their description of a given quantum state due to so-called Unruh effect, the concept of measurement becomes observer-dependent, which implies a nontrivial relativistic modification to the FGUR. For a noninertial observer, we show that the measurements in general mutually unbiased bases (MUBs) could be distinguished from each other, while they correspond to same uncertainty bound in inertial frame. We extend the analysis to an alternative scenario, where, to prevent the Unruh decoherence, an observer is restricted in a single rigid cavity and undergoes an \emph{nonuniform} acceleration. Remarkably, we show that the uncertainty could be drastically degraded by the nonuniform acceleration of cavity during particular epoch, while the uncertainty bound itself exhibits a periodic evolution with respect to the duration of the acceleration. This phenomenon can be attributed to the entanglement generation between the field modes in single cavity that plays the role of quantum memory \cite{FGUR2}. Except the acceleration-duration time with integer periods, the measurements in different MUBs are also distinguishable by the corresponding uncertainty bounds, similar as in scenario with uniform acceleration.

\section{FGUR for an accelerating observer}
\label{2}

We first explore the FGUR for an observer with uniform acceleration $a$, who performs projective measurements on the quantum state constructed from free field modes. For the noninertial observer traveling in, e.g.,  right Rindler wedge (labeled as I), field modes in left Rindler wedge (labeled as II) are unaccessible, as they are separated by acceleration horizon. Therefore, the information loss associated with the horizon can result in a thermal bath perceived by the observer. From the view of quantum information \cite{RQI1}, this celebrated Unruh effect could induce a nontrivial influence on the quantum entanglement between field modes. 

We consider a free massive fermionic field with mass $m$, satisfying the equation $[i\gamma^\mu(\partial_\mu-\Gamma_\mu)+m]\psi=0$, where $\gamma^\mu$ are Dirac matrices and $\Gamma_\mu$ are spin connection. Working in Rindler coordinates, the fermionic field can be expanded in a basis of Unruh modes, which have sharp Rindler frequency and are purely positive frequency linear combinations of Minkowski modes. The general vacuum state in Unruh-basis is therefore Minkowskian, decomposed as $|0_M\ra=|0_U\ra=\bigotimes_{k}|0_{k,R}\ra\otimes|0_{k,L}\ra$ \cite{UNRUH}. Here, subscripts $R$ and $L$ refer to two kinds of Unruh operators that annihilate the vacuum and can be constructed as \cite{SMA1}
\be
C_{k,R}=\c c_{k,\mbox{\tiny I}}-\s d^\dag_{k,\mbox{\tiny II}}~~~~,~~~~C_{k,L}=\c c_{k,\mbox{\tiny II}}-\s d^\dag_{k,\mbox{\tiny I}}.\label{rev1}
\ee
where $\tan r=e^{-\pi\omega/a}$ with Rindler frequency $\omega=\sqrt{k^2+m^2}$, and the particle and antiparticle operators $c_{k,i}$ and $d_{k,i}$ in respective Rindler wedge $\{i=\mbox{I},\mbox{II}\}$ satisfy standard anticommutation relations. The most general annihilation operator in Unruh-basis is therefore a combination of $C_{k,L}$ and $C_{k,R}$ as
\be
C_{k,U}=q_RC_{k,R}^\dag+q_LC_{k,L},\qquad q_R^2+q_L^2=1\label{rev2}
\ee
where $q_R$ and $q_L$ are real parameters. By analytic construction to whole spacetime, the proper Unruh modes are symmetric between Rindler wedges I and II. For particular Rindler frequency $\Omega$, from (\ref{rev1}) and (\ref{rev2}), one can obtain
\bea
|0_{\Omega,R}\ra&=&\c|0_{\Omega,\mbox{\tiny I}}\ra^+|0_{\Omega,\mbox{\tiny II}}\ra^-+\s|1_{\Omega,\mbox{\tiny I}}\ra^+|1_{\Omega,\mbox{\tiny II}}\ra^-\no\\
|0_{\Omega,L}\ra&=&\c|0_{\Omega,\mbox{\tiny I}}\ra^-|0_{\Omega,\mbox{\tiny II}}\ra^+-\s|1_{\Omega,\mbox{\tiny I}}\ra^-|1_{\Omega,\mbox{\tiny II}}\ra^+\no\\
\label{rev3}
\eea
where particle and anti-particle vacua are denoted by $|0\ra^+$ and $|0\ra^-$, similar for excited states. 

The Unruh vacuum therefore becomes \cite{SMA1}
\bea
|0_{\Omega,U}\ra=\cc|0000\ra-\ss|1111\ra+\s\c(|1100\ra-|0011\ra)
\label{unruh-f-0}
\eea 
and the first excitation is 
\bea
|1_{\Omega,U}\ra=q_R(\c|1000\ra-\s|1011\ra)+q_L(\s|1101\ra+\c|0001\ra)
\label{unruh-f-one}
\eea
where we introduce the notations
\be
|1111\ra=b^\dag_{\mbox{\tiny I}}c^\dag_{\mbox{\tiny II}}c^\dag_{\mbox{\tiny I}}b^\dag_{\mbox{\tiny II}}|0_{\Omega,\mbox{\tiny I}}\ra^+|0_{\Omega,\mbox{\tiny II}}\ra^-|0_{\Omega,\mbox{\tiny I}}\ra^-|0_{\Omega,\mbox{\tiny II}}\ra^+
\label{notation}
\ee
It should be noted that different operator ordering in fermonic systems could lead to nonunique results in quantum information \cite{SMA2}. For instance, if we rearrange operator ordering in (\ref{notation}) as $b^\dag_{\mbox{\tiny I}}c^\dag_{\mbox{\tiny I}}c^\dag_{\mbox{\tiny II}}b^\dag_{\mbox{\tiny II}}$, then the Fock basis is changed to $|1111\ra'=-|1111\ra$. In particular, we adopt this so-called physical ordering \cite{SMA3}, as all region I operators appear to the left of all region II operators, which was proposed to guarantee the entanglement behavior of above states would yield physical results. .

To explore how the relativistic motion of observer could influence the FGUR, we consider a scenario in which the state to be measured is prepared in an inertial frame. The observer undergoes an uniform acceleration $a$ can perform measurements on such state in Rindler wedge I, which now should be described in corresponding Rindler frame. As information would loose via Unruh effect, we can expect that the uncertainty obtained by the accelerated observer would be motion-dependent.

We illustrate above intuition by a complete measurements consisting of Pauli operators $\{\sigma_i|i=x,y,z\}$. In particular, we select $\sigma_x$ and $\sigma_z$, behaving as the best measurement basis, where Pauli operators $\sigma_x$ and $\sigma_z$ with equal probability $1/2$ are chosen \cite{FGUR}. Remarkably, along with $\sigma_y$, three set of their eigenvectors form the MUBs in Hilbert space with dimension $d=2$, which plays a central role to theoretical  and practical exploitations of complementarity properties \cite{MUBs}. 
 
For arbitrary pure states $|\psi\ra=\cos\frac{\theta}{2}|0\ra+e^{i\phi}\sin\frac{\theta}{2}|1\ra$ with $\theta\in[0,\pi],\phi\in[0,2\pi)$, the corresponding density matrix $\rho=|\psi\ra\la\psi|$ should be rewritten according to the transformation (\ref{unruh-f-0}) and (\ref{unruh-f-one}). Since the field modes in Rindler wedge II is unaccessible to observer, after tracing over the modes in wedge II, the reduced density matrix $\rho_{red}\equiv\mbox{Tr}_{\mbox{\tiny II}}|\psi\ra\la\psi|$ becomes
\bea
\rho_{red}&=&|00\ra_{\mbox{\tiny I}}\la00|(\cos^2\frac{\theta}{2}c^4+\sin^2\frac{\theta}{2}q_L^2c^2)+|11\ra_{\mbox{\tiny I}}\la11|(\cos^2\frac{\theta}{2}s^4+\sin^2\frac{\theta}{2}q_R^2s^2)\no\\
&+&|01\ra_{\mbox{\tiny I}}\la01|\cos^2\frac{\theta}{2}s^2c^2-|00\ra_{\mbox{\tiny I}}\la11|\sin^2\frac{\theta}{2}q_Rq_Lsc+|10\ra_{\mbox{\tiny I}}\la10|\Big[\cos^2\frac{\theta}{2}s^2c^2+\sin^2\frac{\theta}{2}(q_R^2c^2+q_L^2s^2)\Big]\no\\
&+&\frac{e^{i\phi}}{2}\sin\theta q_Ls\Big(|10\ra_{\mbox{\tiny I}}\la11|s^2-|00\ra_{\mbox{\tiny I}}\la01|c^2\Big)+\frac{e^{-i\phi}}{2}\sin\theta q_Rc\Big(|00\ra_{\mbox{\tiny I}}\la10|c^2+|01\ra_{\mbox{\tiny I}}\la11|s^2\Big)+(h.c.)_{\mbox{\tiny nondiag}}
\eea
with abbreviation $c\equiv\cos r,\ s\equiv\sin r$. After performing the projective measurements $\sigma_x$ and $\sigma_z$ on particle sector, we have the probabilities for the outcomes $(0^x,0^z)$
\bea
p(0^z|\sigma_z)_{\rho}&\equiv&\mbox{tr}(|0\ra^+_{\mbox{\tiny I}}\la0|\rho_{red})=c^2(\cos^2\frac{\theta}{2}+\sin^2\frac{\theta}{2}q_L^2),\no\\
p(0^x|\sigma_x)_\rho&\equiv&\mbox{tr}(|+\ra^+_{\mbox{\tiny I}}\la+|\rho_{red})=\frac{1}{2}+\cos\frac{\theta}{2}\sin\frac{\theta}{2}\cos\phi q_Rc.
\eea
where we chose the measurements in basis $\{|+\ra_{\mbox{\tiny I}},|-\ra_{\mbox{\tiny I}}\}$ and $\{|0\ra_{\mbox{\tiny I}},|1\ra_{\mbox{\tiny I}}\}$, the eigenstates of Pauli matrix $\sigma_x$ and $\sigma_z$ restricted in Rindler wedge I. Therefore, l.h.s. of (\ref{FGUR}) becomes
\bea
U&\equiv&\frac{1}{2}[p(0^z|\sigma_z)_\rho+p(0^x|\sigma_x)_\rho]\no\\
&=&\frac{1}{4}[\sin\theta\cos\phi q_Rc+(\cos\theta q_R^2+q_L^2+1)c^2+1]
\eea
For a particular Unruh mode with fixed acceleration, we estimate the uncertainty bound $\zeta\equiv \mbox{max}_\rho U$ and find that the maximum can always be achieved with $\theta=\frac{\pi}{4}$ and $\phi=0$ \cite{FAN}, i.e., the MCS saturating (\ref{FGUR}) is independent of the acceleration of the observer. This indicates that once we choose the bases enabling an optimal uncertainty in an inertial frame, the corresponding measurements should maintain their optimality for all noninertial observer. This could be useful in many real quantum process, for instance, the BB84 states $\{|+\ra,|-\ra\}$ and $\{|0\ra,|1\ra\}$ in quantum cryptography \cite{BB84}. Hereafter, we always adopt those MCS with $\theta=\frac{\pi}{4}$ and $\phi=0$, and explore the motion-dependence of uncertainty bound $\zeta$ in (\ref{FGUR}) for them.

Explicitly, the dependence of the fine-grained uncertainty bound with outcomes $(0^x,0^z)$ on the acceleration parameter and choice of Unruh modes could be expressed as
\be
\zeta_{(0^x,0^z)}=\frac{1}{4}[c^2(1+q_L^2+\frac{\sqrt{2}}{2} q_R^2)+\frac{\sqrt{2}}{2}q_Rc+1]
\label{bound1}
\ee

Above calculation can extend to any other pairs of outcomes $(0^x,1^z)$, $(1^x,0^z)$ and $(1^x,1^z)$, which all give the same bound $\zeta=\frac{1}{2}+\frac{1}{2\sqrt2}$ in inertial frame \cite{FGUR}. However, we find that the nontrivial Unruh effect could distinguish these four pairs of measurements into two categories. For instance, we have
\bea
p(1^z|\sigma_z)_\rho&\equiv&\mbox{tr}(|1\ra_{\mbox{\tiny I}}\la1|\rho_{red})=\cos^2\frac{\theta}{2}s^2+\sin^2\frac{\theta}{2}(q_R^2+q_L^2s^2),\no\\
p(1^x|\sigma_x)_\rho&\equiv&\mbox{tr}(|-\ra_{\mbox{\tiny I}}\la-|\rho_{red})=\frac{1}{2}-\cos\frac{\theta}{2}\sin\frac{\theta}{2}\cos\phi q_Rc.
\eea
which give
\be
\zeta_{(1^x,1^z)}=\frac{1}{4}[(1+\frac{\sqrt2}{2}c^2)q_R^2+(1+q_L^2)s^2+\frac{\sqrt{2}}{2}q_Rc+1]
\label{bound2}
\ee
for the MCS. By straightforward calculations, it is easy to show that 
\be
\zeta_{(0^x,0^z)}=\zeta_{(1^x,0^z)},~~~~\zeta_{(1^x,1^z)}=\zeta_{(0^x,1^z)}\label{bound+}
\ee

\begin{figure}[hbtp]
\includegraphics[width=.48\textwidth]{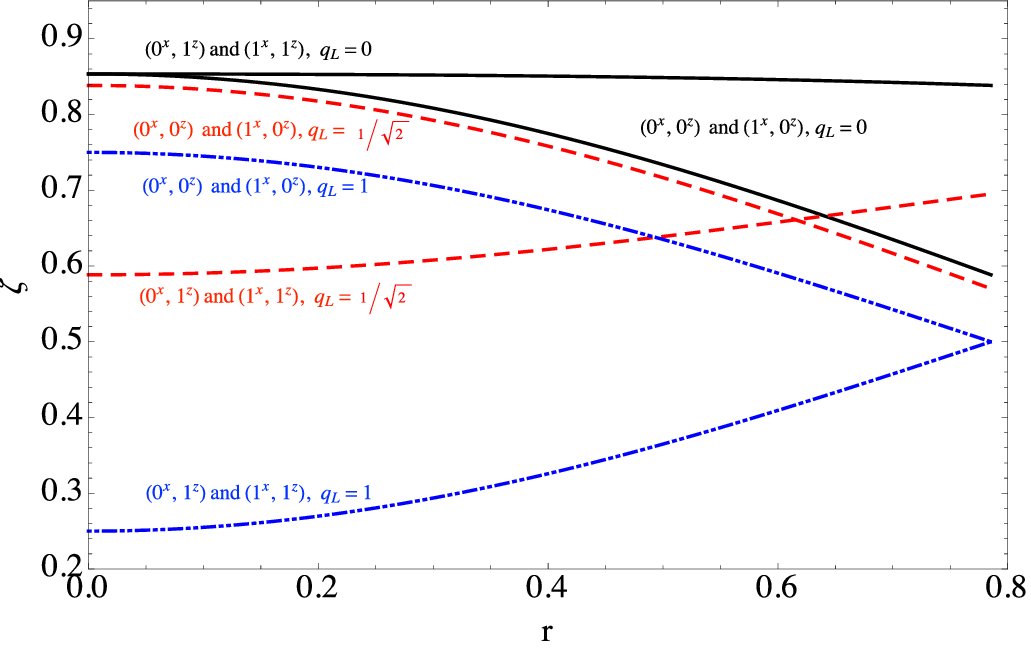}
\caption{The value of $\zeta$ is dependent on the acceleration parameter $r$ and choice of Unruh modes. Three set of curves correspond to the choice of Unruh modes with $q_L=0$ (black solid), $q_L=1/\sqrt{2}$ (red dashed) and $q_L=1$ (blue dashed-double-dotted). Beside the intrinsic quantum uncertainty, Unruh decoherence induces inevitable thermal noise in quantum measurements.}
\label{FGUR1}
\end{figure}

In Fig. \ref{FGUR1}, we depict the uncertainty bounds (\ref{bound1}), (\ref{bound2}) and (\ref{bound+}) for three different choice of Unruh modes. For the case with $q_L=0$, where the Minkowskian annihilation operator is taken to be one of the right or left moving Unruh modes, the noninertial observer would detect a single-mode state once the field is in a special superposition of Minkowski monochromatic modes from an inertial perspective \cite{SMA4}. Under this single-mode approximation (SMA), commonly assumed in the old literature on relativistic quantum information \cite{RQI1}, we recover the standard result $\zeta=\frac{1}{2}+\frac{1}{2\sqrt2}$ for vanishing acceleration. As $r$ growing, the value of $\zeta$ decreases, indicating an increment on measurement uncertainty. This is not surprising as the thermality from inevitable Unruh decoherence introduces classical noise in quantum measurement. In this meaning, the uncertainty bound given in (\ref{FGUR}) quantifies the total uncertainty involved in measurements. On the other hand, for general Unruh modes with $q_L\neq0$, we observe a drastically increase of $\zeta$ w.r.t. growing acceleration for specific measurement outcomes $(1^x,1^z)$ and $(0^x,1^z)$, which means a degradation on measurement uncertainty. Nevertheless, as $\zeta<\frac{1}{2}+\frac{1}{2\sqrt2}$ for infinite acceleration, no violation of standard uncertainty principle can be found in the relativistic scenario with global field modes. 

As illustrated in Fig. \ref{FGUR1}, we find that the distinguishability between the measurements in MUBs is a common feature for any choice of Unruh modes. To explain this, recall that, by definition, a set of orthonormal bases $\{\mathcal{B}_k\}$ for a Hilbert space $\mathcal{H}=\mathbb{C}^d$ where $\mathcal{B}_k=\{|i_k\ra\}=\{|0_k\ra,\cdots,|d-1_k\ra\}$ is called unbiased iff $|\la i_k|j_l\ra|^2=\frac{1}{d}$, $\forall\ i,j\in\{0,\cdots,d-1\}$ holds for all basis vectors $|i_k\ra$ and $|j_l\ra$ with $\forall\ k\neq l$. From an inertial perspective, the MUBs are intimately related to complementarity principle \cite{MUB}, which indicates that the measurement of a observable reveals no information about the outcome of another one if their corresponding bases are mutually unbiased. However, for a noninertial observer,  the bases $\{\mathcal{B}_k\}$ should be transformed according to proper Bogoliubov transformations, which in general breaks the orthonormality. In other word, the MUBs in inertial frame would become non-MUBs from a noninertial perspective. Therefore, for the observer undergoing an uniform acceleration, we expect that Unruh effect could distinguish measurements in MUBs w.r.t. inertial observer.


\section{FGUR for a nonuniform-moving cavity}

We now discuss an alternative scenario in which observer is localized in a rigid cavity, which is more flexible for implement practical quantum information tasks. While the rigid boundaries of the cavity protect the inside observer from the Unruh effect, the relativistic motion of the cavity would still affect the entanglement between the free field modes inside \cite{BOX1,BOX2,BOX3}, therefore leading to a motion-dependent uncertainty bound \cite{FENG2}.

We consider a $(1+1)$-dimensional model, where massless fermionic field is constrained in a cavity with length $L=x_2-x_1$, imposing Dirichlet conditions on the eigenfunctions $\psi_n(t,x)$ of the Hamiltonian. Once the cavity accelerating, it is convenient to use Rindler coordinates $(\eta,\chi)$, defined in the wedge $x>|t|$ by $t=\chi\sinh\eta$ and $x=\chi\cosh\eta$, where $0<\chi<\infty$ and $-\infty<\eta<+\infty$. The new orthonormal eigenfunctions $\hat{\psi}_n(\eta,\chi)$ can be derived by solving the massless Dirac equation $i\gamma^\mu\partial_\mu\hat{\psi}=0$ in Rindler coordinates. 

A typical trajectory of nonuniform-moving cavity contains three segments as (I$'$) the cavity maintains its inertial status initially, then (II$'$) begins to accelerate at $t=0$, following a Killing vector $\partial_\eta$, and finally (III$'$) the acceleration ends at Rindler time $\eta=\eta_1$. The duration of acceleration measured at the center of the cavity is $\tau_1=\frac{1}{2}(x_1+x_2)\eta_1$. The Dirac field can be expanded in quantized eigenfunctions as $\psi=\sum_{n\geqslant0}a_n\psi_n+\sum_{n\leqslant0}b^\dag_n\psi_n$ in segment I$'$, and similarly be expressed by $\hat{\psi}_n$ in segment II$'$ and by $\tilde{\psi}_n$ in segment III$'$. The anticommutators $\{c_m,c^\dag_n\}=\{d_m,d_n^\dag\}=\delta_{mn}$ define the vacuum $c_n|0\ra=d_n|0\ra=0$. Any two field modes in distinct regions can be related by Bogoliubov transformations like $\hat{\psi}_m=\sum_{n}A_{mn}\psi_n$ and $\tilde{\psi}_m=\sum_{n}\mathbb{A}_{mn}\psi_n$, where the coefficients can be calculated perturbatively in the limit of small cavity acceleration \cite{BOX1}. More specifically, by introducing the dimensionless parameter $h=2L/(x_1+x_2)$, a product of cavity's length and acceleration at the center of the cavity, the coefficients can be expanded in a Maclaurin series to $h^2$ order, $A=A^{(0)}+A^{(1)}+A^{(2)}+\mathcal{O}(h^3)$, and similarly $\mathbb{A}=\mathbb{A}^{(0)}+\mathbb{A}^{(1)}+\mathbb{A}^{(2)}+\mathcal{O}(h^3)$.

We start from a pure state $|\psi_k\ra=\cos\frac{\theta}{2}|0_k\ra+e^{i\phi}\sin\frac{\theta}{2}|1_k\ra^+$ in segment I$'$. After an uniform acceleration, we can express this state in segment III$'$ via Bogoliubov transformations, which contains modes within all frequency. Throughout the process, we assume that the observer can only be sensitive to particular modes within frequency $k$. Therefore all other modes with frequency $k'\neq k$ should be traced out in the density matrix $\tilde{\rho}=|\tilde{\psi}_k\ra\la\tilde{\psi}_k|$, which leads to
\bea
\tilde{\rho}_{red}&=&\mbox{tr}_{\neg k}|\tilde{\psi}_k\ra\la\tilde{\psi}_k|\no\\
&=&|\tilde{0}_k\ra\la\tilde{0}_k|(\ctt-\ctt f^-_k+\stt f^+_k)+|\tilde{1}_k\ra^{++}\la\tilde{1}_k|(\stt+\ctt f^-_k-\stt f^+_k)\no\\
&+&|\tilde{0}_k\ra^+\la\tilde{1}_k|\frac{1}{2}\sin\theta e^{-i\phi}(G_k+\mathbb{A}^{(2)}_{kk})+|\tilde{1}_k\ra^+\la\tilde{0}_k|\frac{1}{2}\sin\theta e^{i\phi}(G_k+\mathbb{A}^{(2)}_{kk})^*
\eea
where the coefficients are $f^+_k\equiv\sum_{p\geqslant0}|\mathbb{A}^{(1)}_{pk}|^2$ and $f^-_k\equiv\sum_{p<0}|\mathbb{A}^{(1)}_{pk}|^2$. The probability of measurements $(0^x,0^z)$ are
\bea
p(0^z|\sigma_z)_{\tilde{\rho}}&\equiv&\mbox{tr}(|\tilde{0}_k\ra\la\tilde{0}_k|\tilde{\rho}_{red})=\ctt-\ctt f^-_k+\stt f^+_k,\no\\
p(0^x|\sigma_x)_{\tilde{\rho}}&\equiv&\mbox{tr}(|+\ra\la+|\tilde{\rho}_{red})=\frac{1}{2}\{1+\sin\theta\mbox{Re}[e^{-i\phi}(G_k+\mathbb{A}^{(2)}_{kk})]\}
\eea
The uncertainty bound should be the maximum of l.h.s. of (\ref{FGUR}), giving $U\equiv\frac{1}{2}[p(0^z|\sigma_z)_{\tilde{\rho}}+p(0^x|\sigma_x)_{\tilde{\rho}}]$. Along a similar analysis as before, we know that the acceleration of cavity would not change the MCS with parameters $\theta=\frac{\pi}{4}$ and $\phi=0$. Therefore, we obtain the uncertainty bound for the cavity system
\be
\tilde{\zeta}_{(0^x,0^z)}=\frac{1}{4\sqrt2}[1+2\sqrt2-F_++\sqrt2F_-+\mbox{Re}(G_k+\mathbb{A}^{(2)}_{kk})]
\label{fgur3}
\ee
The coefficients in the bound has been given in \cite{BOX1}, which are
\bea
&&F_+=\sum_{p=-\infty}^{\infty}|E_1^{k-p}-1|^2|A^{(1)}_{kp}|^2=\frac{4h^2}{\pi^4}[4(k+s)^2(Q_6(1)-Q_6(E_1))+Q_4(1)-Q_4(E_1)]\no
\eea
and \cite{FENG2}
\bea
&&F_-\equiv f^+_k-f^-_k=\bigg(\sum_{p\geqslant0}-\sum_{p<0}\bigg)|E_1^{k-p}-1|^2|A^{(1)}_{kp}|^2=\frac{16h^2}{\pi^4}2(k+s)[Q_5(1)-Q_5(E_1)]+P(k,s,E_1)\no
\eea
 with $s\in[0,1)$ characterizing the self-adjoint extension of the Hamiltonian. Here we use the notation $Q_\alpha(\beta)\equiv\mbox{Re}\big[\mbox{Li}_\alpha(\beta)-\frac{1}{2^\alpha}\mbox{Li}_\alpha(\beta^2)\big]$, Li is the polylogarithm and $E_1\equiv\exp(\frac{i\pi\eta_1}{\ln(x_2/x_1)})=\exp(\frac{i\pi h\tau_1}{2Lx_1\tanh(h/2)})$. $P$ is a polynomial summing for all terms with odd number
\be
\sum_{\substack{odd\\   m=1}}^{k}\frac{4h^2}{\pi^4}\big(1-\mbox{Re}(E_1^{m})\big)\bigg[4(k+s)\bigg(\frac{k+s}{m}-1\bigg)+\frac{1}{m^4}\bigg]
\no
\ee
and 
\bea
\mbox{Re}(G_k+\mathbb{A}^{(2)}_{kk})&=&1-h^2\bigg\{\bigg(\frac{1}{48}+\frac{\pi^2(k+s)^2}{120}\bigg)-\frac{2}{\pi^4}\big[4(k+s)^2Q_6(E_1)+Q_4(E_1)\big]\bigg\}\no
\eea

In previous section, we show that an uniformly-accelerating observer can distinguish the measurements in MUBs which share the same uncertainty bounds in an inertial frame. Here we generalize this to the scenario with rigid cavity. To proceed, we calculate the probabilities of measurements $(1^x,1^z)$
\bea
p(1^z|\sigma_z)_{\tilde{\rho}}&\equiv&\mbox{tr}(|\tilde{1}_k\ra^{++}\la\tilde{1}_k|\tilde{\rho}_{red})=\stt+\ctt f^-_k-\stt f^+_k,\no\\
p(1^x|\sigma_x)_{\tilde{\rho}}&\equiv&\mbox{tr}(|+\ra\la+|\tilde{\rho}_{red})=\frac{1}{2}\{1-\sin\theta\mbox{Re}[e^{-i\phi}(G_k+\mathbb{A}^{(2)}_{kk})]\}
\eea
which give the uncertainty bound $\tilde{\zeta}_{(1^x,1^z)}$ for MCS
\be
\tilde{\zeta}_{(1^x,1^z)}=\frac{1}{4\sqrt2}[1+2\sqrt2-F_+-\sqrt2F_-+\mbox{Re}(G_k+\mathbb{A}^{(2)}_{kk})]\label{fgur4}
\ee
Similar as before, by straightforward calculations, we can further prove that 
\be
\tilde{\zeta}_{(0^x,0^z)}=\tilde{\zeta}_{(1^x,0^z)},~~~~\tilde{\zeta}_{(1^x,1^z)}=\tilde{\zeta}_{(0^x,1^z)}\label{bound++}
\ee

\begin{figure}[hbtp]
\includegraphics[width=.48\textwidth]{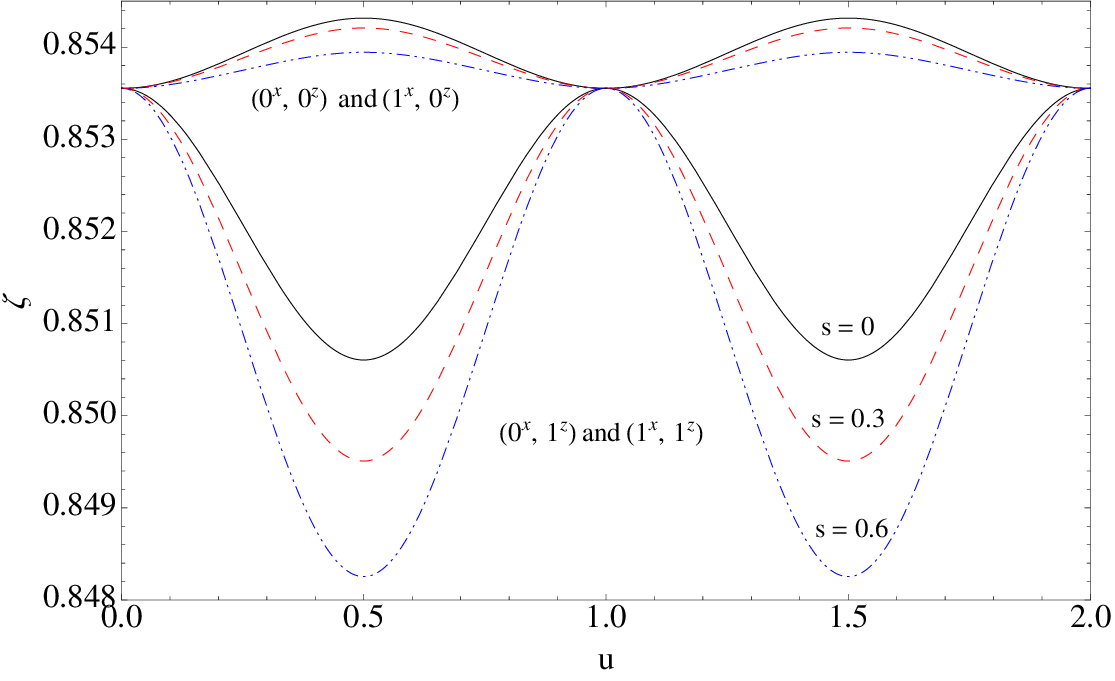}
\caption{The value of $\zeta$ depends on the duration time of acceleration of rigid cavity. We choose $k=1$. For each pair of measurements ($(0^x,0^z)$ and $(1^x,0^z)$, $(1^x,1^z)$ and $(0^x,1^z)$), three cures from top to bottom correspond to parameters $s=0,\ 0.3,\ 0.6$. The parameter $u=\eta_1/(2\ln(x_2/x_1))=h\tau_1/[4Lx_1\tanh{h/2}]$ characterizes the duration time of cavity acceleration. To demonstrate the low acceleration approximation, the uncertainty is estimated under $h=0.1$.}
\label{FGUR3}
\end{figure}

We depict above uncertainty bounds for measurements performed within cavity in Fig. \ref{FGUR3}. Firstly, we find that both uncertainty bounds (\ref{fgur3}), (\ref{fgur4}) and (\ref{bound++}) are now periodic in time $\tau_1$, which measures the duration of the cavity acceleration, with the period $T=4Lx_1\tanh(h/2)/h$. By properly choosing the parameters to ensure that $\tau_1=nT$ with $n\in \mathbb{N}$, the uncertainty bounds are protected \cite{FENG2}, recovering the value $\frac{1}{2}+\frac{1}{2\sqrt2}$ as in inertial case. However, we observe an interesting violation of uncertainty bound $\frac{1}{2}+\frac{1}{2\sqrt2}$ for quantum measurements with specific outcomes, such as $(0^x,0^z)$ and $(1^x,0^z)$. This can be interpreted as a result of the entanglement generation between the field modes in the single rigid cavity that plays the role of quantum memory \cite{FGUR2}. Therefore, with employing the relativistic effect on localized quantum system, one may achieve lower uncertainty bound and higher precision of outcomes prediction than in the non-relativistic frame. 

For arbitrary acceleration duration $\tau_1\neq nT$,  measurements with outcomes $(0^x,0^z)$ and $(1^x,1^z)$ can be distinguished from each other by the corresponding uncertainty bounds (\ref{fgur3}) and (\ref{fgur4}). Therefore, same as first scenario with Unruh effect, we conclude that the relativistic motion of a rigid cavity can provoke the distinguishability between the measurements in MUBs that share the same bound $\frac{1}{2}+\frac{1}{2\sqrt2}$ for an inertial observer. 

Finally, we would like to remark the experimental side of our results. While most of experimental tests of relativistic quantum information require an extremely high acceleration for global field modes, in localized scenario, the only approximation in our analysis was to work in the small acceleration regime with $|k|h\ll1$. Numerical estimation suggests \cite{BOX2} that the magnitude of relativistic modification already become observable at microgravity acceleration of $10^{-10}$ $\mbox{ms}^{-2}$. Indeed, with high-precision quantum optics techniques, the upper bound of quantum entanglement in non-inertial reference frames has been determined recently \cite{ex1} to uniform acceleration imposed by low-$g$ free-fall motion. As the range of our estimation is also under the ability of such cutting-edge technology, following above lines \cite{ex2}, we believe that a novel cavity experiment in non-inertial frame could be proposed in near future to test the relativistic modification of quantum information protocol.

\section{Discussions}
\label{4}

In this Letter, we explored the nontrivial relativistic modification to the FGUR. We have shown that, for an observer undergoing a large acceleration, the associated Unruh effect could increase or reduce the fine-grained uncertainty bounds, depending on the choice of Unruh modes. Moreover, we have shown that the measurements in MUBs, sharing same uncertainty bound in inertial frame, could be distinguished from each other when the observer undergoes a nonvanishing acceleration. In an alternative scenario, we have investigated the FGUR for the measurements on fermionic field modes restricted in a single rigid cavity, where the uncertainty  bound itself exhibits a periodic evolution w.r.t. the duration of the acceleration. For quantum measurements with specific outcomes, we find an interesting violation of uncertainty bound in inertial frame, attributed to the entanglement generation in cavity that plays the role of quantum memory. Our results provide a novel way to investigate the relativistic effect in a quantum-information context, which may be experimental tested by future quantum metrology \cite{METROLOGY}.

Throughout the Letter, we only discuss the influence of Unruh effect on FGUR for fermionic field, which we believe that manifest more abundant characteristics of motion-dependence of FGUR in relativistic framework than bosonic field. For instance, while there is no fundamental difference in Unruh decoherence for different choice of bosonic Unruh modes \cite{SMA1}, it implies that measurement uncertainty on bosonic states should always increase with growing acceleration of observer. In this meaning, no bosonic Unruh mode can degrade measurement uncertainty like fermionic Unruh mode with special $q_L\neq0$ does in Fig.\ref{FGUR1}. On the other hand, in a cavity scenario, Unruh decoherence on localized bosonic modes exhibits a periodic-dependence on the duration of acceleration, similar as fermionic case \cite{BOX2}. Therefore, one could expect that for bosonic field, comparing to Fig.\ref{FGUR3}, no essential change but only numerical difference on the FGUR bound $\xi$ would happen.

Our study raises several implications. Firstly, we can generalize above analysis to fundamental MUBs in higher dimensional Hilbert spaces \cite{MUB4,MUB5}, where a $n$-qubit can be truncated from free scalar field modes with infinite levels. On the other hand, the fascinating link between FGUR and the second thermodynamical law has been explored in \cite{FGUR3}, which proved that a deviation of the FGUR implies a violation of the second law of thermodynamics. In this spirit, by investigating the influence of the relativistic motion of observer on a thermodynamical cycle, one could relate the relativistic effect to thermodynamics in an information-theoretic way. Finally, we can explore the FGUR in some dynamical spacetimes \cite{FENG3}, e.g., cosmological background, where the entanglement generated through the evolution of spacetime is expected to play a significant role in quantum measurements \cite{COS}.

\section*{ACKNOWLEDGEMENT}
This work is supported by the National Natural Science Foundation
of China (No. 11505133), the Fundamental Research Funds for the Central Universities, Natural Science Basic Research Plan in Shaanxi Province of China (No. 2018JM1049) and the Postdoctoral Science
Foundation of China (No. 2016M592769). H.F. acknowledges the 
support of the National Natural Science Foundation of China (No. 91536108, 11774406) and National Key R\&D Program of China (No. 2016YFA0302104, 2016YFA0300600). 
Y.Z.Z. acknowledges the support of the
Australian Research Council Discovery Project (No. 140101492) and the National Natural Science Foundation
of China (No. 11775177).

\end{document}